\documentclass[12pt,draftclsnofoot,onecolumn,romanappendices]{IEEEtran}

\IEEEoverridecommandlockouts

\usepackage{algorithm,algorithmic,amsmath,amssymb,amsthm,array,bbm,bibentry,cite,color,comment}
\usepackage{enumerate,enumitem,eurosym,float,graphicx,lettrine,mathrsfs,multirow,nomencl}
\usepackage{pict2e,psfrag,ragged2e,setspace,stfloats,subfigure,supertabular,tabularx,url}
\usepackage[USenglish]{babel}

{		\newtheorem{theorem}{Theorem}}
{ 		}
{ 		\newtheorem{lemma}{Lemma}}
{ 		\newtheorem{proposition}{Proposition}}
{ 		\newtheorem{remark}{Remark}}
{		}
{ 		\newtheorem{corollary}{Corollary}}

\newcommand{\setL}{\mathcal{L}}

\newcommand{\Real}{\mbox{$\mathbb{R}$}}

\newcommand{\argmax}{\operatornamewithlimits{argmax}}

\newcommand{\diff}{\mathrm{d}}
\newcommand{\Exp}{\mathbb{E}}

\renewcommand{\Pr}{\mathbb{P}}

\newcommand{\sir}{\mathrm{SIR}}
\newcommand{\sinr}{\mathrm{SINR}}
\newcommand{\los}{\textnormal{\tiny{LOS}}}
\newcommand{\nlos}{\textnormal{\tiny{NLOS}}}
\newcommand{\Pcov}{\mathrm{P}_{\mathrm{cov}}}
\newcommand{\rmQ}{\textnormal{\tiny{Q}}}
\newcommand{\rmC}{\mathrm{C}}
\newcommand{\rmS}{\mathrm{S}}
\newcommand{\ASE}{\mathrm{ASE}}

\title{\huge{Downlink Cellular Network Analysis with LOS/NLOS Propagation and Elevated Base Stations}}
\author{Italo Atzeni,~\IEEEmembership{Member,~IEEE}, Jes\'{u}s~Arnau,~\IEEEmembership{Member,~IEEE}, \\ and Marios Kountouris,~\IEEEmembership{Senior Member,~IEEE}\thanks{Parts of this work were presented at IEEE ICC 2016 in Kuala Lumpur, Malaysia \cite{Arn16}, and is submitted to SpaSWiN 2017 (WiOpt Workshops) in Paris, France \cite{Atz17a}.} \thanks{The authors are with the Mathematical and Algorithmic Sciences Lab, France Research Center, Huawei Technologies France SASU, 92100 Boulogne-Billancourt, France (email: \{italo.atzeni, jesus.arnau, marios.kountouris\}@huawei.com).}}

\begin{document}

\maketitle

\begin{abstract}
In this paper, we investigate the downlink performance of dense cellular networks with elevated base stations (BSs) using a channel model that incorporates line-of-sight (LOS)/non-line-of-sight (NLOS) propagation in both small-scale and large-scale fading. Modeling LOS fading with Nakagami-$m$ fading, we provide a unified framework based on stochastic geometry that encompasses both closest and strongest BS association. Our study is particularized to two distance-dependent LOS/NLOS models of practical interest. Considering the effect of LOS propagation alone, we derive closed-form expressions for the coverage probability with Nakagami-$m$ fading, showing that the performance for strongest BS association is the same as in the case of Rayleigh fading, whereas for closest BS association it monotonically increases with the shape parameter $m$. Then, focusing on the effect of elevated BSs, we show that network densification eventually leads to near-universal outage even for moderately low BS densities: in particular, the maximum area spectral efficiency is proportional to the inverse of the squared BS height.

\end{abstract}


\section{Introduction} \label{sec:intro}


Ultra-dense networks (UDNs), i.e., dense and massive deployments of small-cell base stations (BSs) with wired/wireless backhaul connectivity, are foreseen as a core element to realize the vision of 5th generation (5G) wireless systems. UDNs are expected to achieve higher data rates and enhanced coverage by exploiting spatial reuse while retaining seamless connectivity and low energy consumption \cite{Bhu14,Que13}. Recent studies using stochastic geometry models have shown that the throughput grows linearly with the BS density in the absence of background noise and for closest BS association \cite{And11}, i.e., when each user equipment (UE) is associated with the closest BS; similar results are reported in \cite{Dhi12} for strongest BS association, i.e., when each UE is associated with the BS with the highest signal-to-interference-plus-noise ratio (SINR). Nevertheless, most prior performance analyses assume simple models--mostly for tractability reasons--in which \textit{i)} BSs are located according to a homogeneous Poisson point process (PPP) and are placed at the same height as the UEs, and \textit{ii)} the signal propagation is modeled using the standard single-slope pathloss and the Rayleigh distribution for the small-scale fading.

In parallel with UDNs coming to prominence, there has been a growing interest in devising increasingly realistic models for their system-level performance evaluation. In this respect, \cite{Zha15} studies the impact of dual-slope pathloss on the performance of downlink UDNs and shows that both coverage and throughput strongly depend on the network density. In \cite{Bai15}, a stochastic geometry based framework for millimeter wave and pathloss with line-of-sight (LOS) and non-line-of-sight (NLOS) propagation is proposed. More comprehensive models can be found in \cite{Gal15,Din16,Gup16}, where the pathloss exponent changes with a probability that depends on the distance between BSs and UEs. Under such models, the throughput does not necessarily grow monotonically with the BS density due to the different scaling of desired signal and interference; this is further investigated in \cite{Din16b}, where the effect of shutting down idle BSs is taken into account. Moreover, coverage and rate scaling laws in UDNs using regular variation theory are derived in \cite{Ngu16}. Lastly, \cite{DiR16} introduces an approximation that allows to obtain simpler expressions with the above mentioned models while incorporating blockage effects and non-isotropic antenna patterns.

Several previous works effectively capture the effect of LOS propagation on the large-scale fading (i.e., the pathloss), although they do so by modeling the small-scale fading using the Rayleigh distribution. This assumption is justified for NLOS propagation and is widely adopted mainly due to tractability. However, when the BS density increases, assuming NLOS for the received signal may not be realistic, significantly altering the coverage and throughput performance; in fact, it is commonly accepted that LOS propagation is subject to Rician fading. In addition to modifying the pathloss exponent according to a distance-dependent probability function, \cite{Arn16,And16} consider varying the small-scale fading distribution as well. Likewise, previous studies often neglect the possible difference in height between BSs and UEs, which implicitly sets a limit on how close UEs can be to their serving BSs regardless of the BS density. The practically relevant case of elevated BSs has been recently incorporated into the discussion on network densification by our previous work \cite{Atz17a} and \cite{Din16a}, appeared during the preparation of this manuscript, revealing that the elevation difference between BSs and UEs has a detrimental effect on the system performance.

\subsection{Contributions} \label{sec:intro_contr}

The overall contribution of this paper broadens prior studies on network densification by investigating the downlink performance of UDNs using a model that incorporates elevated BSs and dual-slope LOS/NLOS propagation effects in both small-scale and large-scale fading. More specifically, the main contributions are as follows:
\begin{itemize}
\item[$\bullet$] Modeling Rician fading by means of Nakagami-$m$ fading, we propose a general framework based on stochastic geometry that accommodates both closest and strongest BS association; as performance metrics, we consider the coverage probability and the area spectral efficiency (ASE). In addition, we particularize our study to two practical distance-dependent LOS/NLOS models, i.e., the widely used 3GPP model and a newly proposed model with randomly placed buildings.
\item[$\bullet$] Considering the effect of LOS propagation alone, we derive closed-form expressions of the coverage probability with Nakagami-$m$ fading and provide useful asymptotic trends. Interestingly, the coverage probability for strongest BS association is the same as in the case of Rayleigh fading, whereas for closest BS association it monotonically increases with the shape parameter $m$ until it converges to the value obtained for strongest BS association. In general, the performance turns out to be dominated by the pathloss and is marginally affected by the small-scale fading.
\item[$\bullet$] Considering the effect of elevated BSs alone, we characterize the interference power and derive both closed-form (for closest BS association) and integral (for strongest BS association) expressions of the coverage probability and the optimal BS density. In particular, we show that the maximum ASE is proportional to the inverse of the squared BS height. Indeed, the BS height proves to be the most prominent factor in degrading the system performance, since it leads to near-universal outage regardless of the other parameters and even at moderately low BS densities.
\end{itemize}

The remainder of the paper is structured as follows. Section~\ref{sec:SM} introduces the system model. Section~\ref{sec:cov} provides expressions for the coverage probability using general distance-dependent LOS/NLOS models. Section~\ref{sec:los} analyzes the effect of Nakagami-$m$ fading, whereas Section~\ref{sec:height} focuses on the impact of BS height. In Section~\ref{sec:num}, numerical results are reported to corroborate our theoretical findings and to quantify the individual and combined effect of each of the above factors. Finally, Section~\ref{sec:concl} summarizes our contributions and draws some concluding remarks.

\section{System Model} \label{sec:SM}
\subsection{Network Model} \label{sec:SM_net}

We consider a dense downlink cellular network, in which the location distribution of the single-antenna BSs\footnote{The case of multi-antenna BSs is considered in Appendix~\ref{sec:A1_MISO}.} is modeled according to a marked PPP $\widehat{\Phi} \triangleq \{(x_{i}, g_{x_{i}})\} \subset \Real^{2} \times \Real^{+}$. The underlying point process $\Phi \triangleq \{ x_{i} \} \subset \Real^{2}$ is a homogeneous PPP with density $\lambda$, measured in [BSs/m$^{2}$], and the mark $g_{x_{i}} \in \Real^{+}$ represents the channel power fading gain from the BS located at $x_{i}$ to a randomly chosen downlink UE referred to as \textit{typical UE}, which is located at the origin of the Euclidean plane. In this setting, the employment of PPPs allows to capture the spatial randomness of real-world UDN deployments (often not fully coordinated) and, at the same time, obtain precise and tractable expressions for system-level performance metrics \cite{And11,Hae12}; considering more involved random spatial models goes beyond the scope of this paper. The UEs, also equipped with a single antenna, are distributed according to some independent and homogeneous point process $\Phi_{\mathrm{u}}$ (e.g., PPP) whose intensity $\lambda_{\mathrm{u}}$ is sufficiently larger than $\lambda$ in order to ensure that each BS is active, i.e., it has at least one UE associated within its coverage. The typical UE is associated with a \textit{serving BS} following one of the BS association policies described in Section~\ref{sec:SM_SINR}; the remaining BSs are thus \textit{interfering BSs}. Lastly, we assume that all BSs are elevated at the same height $h \geq 0$, measured in [m], whereas the typical UE is at the ground level; alternatively, $h$ can be interpreted as the elevation difference between BSs and UEs if the latter are all placed at the same height.

\subsection{Channel Model} \label{sec:SM_ch}

Let $r_{x} \triangleq \| x \|$ denote the horizontal distance between $x$ and the typical UE, measured in [m]. We consider a distance-dependent LOS probability function $p_{\los}(r_{x})$, i.e., the probability that a BS located at $x$ experiences LOS propagation depends on the distance $r_{x}$. Therefore, we use $\Phi_{\los} \triangleq \{ x  \in \Phi : x {\rm \ in \ LOS} \}$ and $\Phi_{\nlos} \triangleq \Phi \, \backslash \, \Phi_{\los}$ to denote the subsets of BSs in LOS and in NLOS propagation conditions, respectively. We remark that each BS is characterized by either LOS or NLOS propagation independently from the others and regardless of its operating mode as serving or interfering BS.

The propagation through the wireless channel is characterized as the combination of pathloss attenuation and small-scale fading. For the former, we adopt the standard power-law pathloss model and define the pathloss functions $\ell_{\los}(r_{x},h) \triangleq (r_{x}^2 + h^2)^{-\frac{\alpha_{\los}}{2}}$ if $x \in \Phi_{\los}$ and $\ell_{\nlos}(r_{x},h) \triangleq (r_{x}^2 + h^2)^{-\frac{\alpha_{\nlos}}{2}}$ if $x \in \Phi_{\nlos}$, with $\alpha_{\nlos} \geq \alpha_{\los} > 2$.  For the latter, we assume that the channel amplitudes are Nakagami-$m$ distributed for LOS propagation conditions and Rayleigh distributed for NLOS propagation conditions. Observe that the commonly used Rician distribution is well approximated by the more tractable Nakagami-$m$ distribution with the shape parameter $m$ computed as $m \triangleq (K+1)^{2}/(2K+1)$, where $K$ is the Rician $K$-factor representing the ratio between the powers of the direct and scattered paths.\footnote{Note that, in order to use such formulation, the value of $m$ is rounded to the closest integer.} Hence, the channel power fading gain $g_{x}$ follows the Gamma distribution $\Gamma \big( m, \frac{1}{m} \big)$ if $x \in \Phi_{\los}$, with complementary cumulative distribution function (CCDF) given by
\begin{align}
\label{eq:ccdf_los} \bar{F}_{\los}(z) & \triangleq 1 - \frac{\gamma(m, m z)}{\Gamma(m)} = e^{-m z} \sum_{k=0}^{m-1} \frac{(m z)^k}{k!}
\end{align}
where the last equality holds when the shape parameter $m$ is an integer; on the other hand, $g_{x}$ follows the exponential distribution $\exp(1)$ if $x \in \Phi_{\nlos}$ and its CCDF $\bar{F}_{\nlos}(z)$ can be obtained from $\bar{F}_{\los}(z)$ in \eqref{eq:ccdf_los} by simply setting $m=1$.

\subsection{SINR and BS Association} \label{sec:SM_SINR}

The SINR when the typical UE is associated to the BS located at $x$ is given by
\begin{align} \label{eq:SINR}
\sinr_{x} \triangleq \frac{g_{x} \ell_{\rmQ}(r_{x},h)}{I + \sigma^{2}}
\end{align}
where the sub-index $\mathrm{Q}$ takes the form $\mathrm{Q} = \mathrm{LOS}$ if $x \in \Phi_{\los}$ and $\mathrm{Q} = \mathrm{NLOS}$ if $x \in \Phi_{\nlos}$, $I$ is the aggregate interference power defined as
\begin{align} \label{eq:I}
I \triangleq \sum_{y \in \Phi_{\los} \backslash \{x\}} g_{y} \ell_{\los}(r_{y},h) + \sum_{y \in \Phi_{\nlos} \backslash \{x\}} g_{y} \ell_{\nlos}(r_{y},h)
\end{align}
and $\sigma^{2}$ is the additive noise power. For the sake of simplicity, we consider the interference-limited case, i.e., $I \gg \sigma^{2}$, and we thus focus on the signal-to-interference ratio (SIR). Our analysis can be extended with more involved calculations to the general case. 

In this paper, we consider a unified framework that encompasses both closest \cite{And11} and strongest (i.e., highest SINR) \cite{Dhi12} BS association. For this purpose, we introduce the following preliminary definitions \cite{Arn16}:
\begin{align}
\label{eq:phi} f_{r_{x}}(r) & \triangleq \left\{
\begin{array}{ll}
2 \pi \lambda e^{- \pi \lambda r^{2}} r, & \quad \textrm{closest BS} \\
2 \pi \lambda r, & \quad \textrm{strongest BS}
\end{array} \right. \\
\label{eq:nu} \nu(r) & \triangleq \left\{
\begin{array}{ll}
r, & \hspace{17.5mm} \quad \textrm{closest BS} \\
0, & \hspace{17.5mm} \quad \textrm{strongest BS}
\end{array} \right.
\end{align}
where $f_{r_{x}}(r)$ in \eqref{eq:phi} represents the probability density function (PDF) of the distance $r_{x}$ between the serving BS and the typical UE.

\section{Coverage Probability} \label{sec:cov}

In this section, we provide the general expression of the coverage probability when both serving and interfering BSs independently experience LOS or NLOS propagation conditions with respect to the typical UE depending on their distance from the latter. The coverage probability is defined as the probability that the received SIR is larger than a target SIR threshold $\theta$, i.e., $\Pcov (\theta) \triangleq \Pr [\sir_{x} > \theta]$. The coverage probability allows to compute the achievable ASE, defined as $\ASE (\theta) \triangleq \lambda \Pcov (\theta) \log_{2} (1 + \theta)$, measured in [bps/Hz/m$^{2}$].

Let us use $\setL_{I}^{\los} (s)$ and $\setL_{I}^{\nlos} (s)$ to denote the Laplace transforms of the interference when $p_\los(r)=1$ and $p_\los(r)=0$, $\forall r \in [0, \infty)$, respectively, which correspond to the cases of LOS or NLOS interference:
\begin{align}
\label{eq:L_I_los} \setL_{I}^{\los} (s) & \triangleq \exp \bigg( - 2 \pi \lambda \int_{\nu(r)}^{\infty} \bigg( 1 - \frac{1}{(1 + \frac{s}{m} \ell_{\los} (t,h))^{m}} \bigg) t \diff t \bigg) \\
\label{eq:L_I_nlos} \setL_{I}^{\nlos} (s) & \triangleq \exp \bigg( - 2 \pi \lambda \int_{\nu(r)}^{\infty} \bigg( 1 - \frac{1}{1 + s \ell_{\nlos} (t,h)} \bigg) t \diff t \bigg).
\end{align}

\subsection{General LOS/NLOS Model} \label{sec:cov_general}

We begin by considering a general expression of $p_\los(r)$. In this setting, the coverage probability is formalized in the following theorem.

\begin{theorem} \label{th:P_cov} \rm{
The coverage probability is given by
\begin{align}
\label{eq:P_cov} \Pcov (\theta) \! = \! \int_{0}^{\infty} \! \bigg( p_{\los}(r) \sum_{k=0}^{m-1} & \bigg[ \frac{(-s)^{k}}{k!} \frac{\diff^{k}}{\diff s^{k}} \setL_{I} (s) \bigg]_{s = m \theta / \ell_{\los}(r,h)} \! \! \! + \big( 1 \! - \! p_{\los}(r) \big) \setL_{I} \bigg( \frac{\theta}{\ell_{\nlos}(r,h)} \bigg) \bigg) f_{r_{x}}(r) \diff r
\end{align}
where
\begin{align}
\label{eq:L_I} \setL_{I} (s) & \triangleq \setL_{I}^{\nlos} (s) \exp \bigg( - 2 \pi \lambda \int_{\nu(r)}^{\infty} p_{\los}(t) \bigg( \frac{1}{1 + s \ell_{\nlos}(t,h)} - \frac{1}{(1 + \frac{s}{m} \ell_{\los}(t,h))^{m}} \bigg) t \diff t \bigg)
\end{align}
is the Laplace transform of the interference $I$ in \eqref{eq:I}, with $\setL_{I}^{\nlos} (s)$ defined in \eqref{eq:L_I_nlos}.}
\end{theorem}

\begin{IEEEproof}
See Appendix~\ref{sec:A1_P_cov}.
\end{IEEEproof} \vspace{1mm}

\noindent The result of Theorem~\ref{th:P_cov} is extended to the case of multi-antenna BSs in Appendix~\ref{sec:A1_MISO}.

\begin{remark} \rm{
Due to the contribution from the interfering BSs in LOS propagation conditions, we have that $\setL_{I} (s) \leq \setL_{I}^{\nlos} (s)$, with $\setL_{I}^{\nlos} (s)$ in \eqref{eq:L_I_nlos}. This can be equivalently seen from the argument of the exponential function in \eqref{eq:L_I}, which is always negative since $\alpha_{\nlos} > \alpha_{\los}$ and $m \geq 1$. On the other hand, the possibility of LOS desired signal enhances the coverage probability in \eqref{eq:P_cov}.}
\end{remark}

\begin{remark} \rm{
Observe that $\setL_{I}(s)$ in \eqref{eq:L_I} reduces to $\setL_{I}^{\los}(s)$ in \eqref{eq:L_I_los} if $p_{\los}(r)=1$, $\forall r \in [0, \infty)$ (see Section~\ref{sec:los}) and to $\setL_{I}^{\nlos}(s)$ in \eqref{eq:L_I_nlos} if $p_{\los}(r)=0$, $\forall r \in [0, \infty)$.}
\end{remark}

\noindent So far, it is not straightforward to get clear insights on how fading, pathloss, and BS height individually affect the network performance. Hence, in Sections~\ref{sec:los} and \ref{sec:height}, we separately examine the effect of Nakagami-$m$ fading and BS height, respectively. 

Let $\Pcov^{\nlos}(\theta)$ (resp. $\Pcov^{\los}(\theta)$) denote the coverage probability in presence of NLOS (resp. LOS) propagation. The following corollary provides the asymptotic trends of the coverage probability in \eqref{eq:P_cov}.

\begin{corollary} \label{cor:3gpp} \rm{
Assume that the LOS probability function $p_{\los}(r)$ is monotonically decreasing with $r$. Then, the following hold:
\begin{itemize}
\item[\textit{(a)}] $\displaystyle \lim_{\lambda \to 0} \mathrm{P}_{\mathrm{cov}}(\theta) = \mathrm{P}_{\mathrm{cov}}^{\nlos}(\theta)$;
\item[\textit{(b)}] $\displaystyle \lim_{\lambda \to \infty} \mathrm{P}_{\mathrm{cov}}(\theta) = \mathrm{P}_{\mathrm{cov}}^{\los}(\theta)$.
\end{itemize}}
\end{corollary}

\begin{IEEEproof}
\textit{(a)} Since the average distance between the typical UE and the $n$th nearest BS is proportional to $\frac{1}{\sqrt{\lambda}}$ \cite[Ch.~2.9]{Hae12}, when $\lambda \to 0$ we have $\Phi_{\los} = \emptyset$ almost surely. \textit{(b)} Likewise, when $\lambda \to \infty$, the serving BS is in LOS propagation conditions almost surely, whereas the interference is also dominated by BSs in LOS propagation conditions.
\end{IEEEproof} \vspace{1mm}

The derivatives of the Laplace transform of the interference arise in presence of multiple signal components as, e.g., when the received signal is subject to Nakagami-$m$ fading (as in \eqref{th:P_cov} above) or when multiple antennas are involved (as in \cite{Atz17,Hun08}). A useful upper bound for this type of expression is provided in the following proposition.\footnote{A lower bound with a similar expression can be also obtained; however, such bound is usually not sufficiently tight and it is thus not considered.}

\begin{proposition} \label{pro:Alzer} \rm{
For any $\setL_{X}(z) \triangleq \Exp_{X} \big[ e^{-z X} \big]$ and $N > 1$, the following inequality holds:
\begin{align}
\label{eq:Alzer} \sum_{n=0}^{N - 1} \bigg[ \frac{(- s)^{n}}{n!} \frac{\mathrm{d}^{n}}{\mathrm{d} s^{n}} \mathcal{L}_{X}(s) \bigg]_{s = z} < \sum_{n=1}^{N} (-1)^{n - 1} {{N}\choose{n}} \setL_{X} \big( n \big( \Gamma(N + 1) \big)^{- \frac{1}{N}} z \big).
\end{align}}
\end{proposition}

\begin{IEEEproof}
The upper bound is based on Alzer's inequality; we refer to \cite{Atz17} for details.
\end{IEEEproof} \vspace{1mm}

So far we have assumed no particular expression for $p_\los(r)$. In Sections~\ref{sec:cov_3gpp} and \ref{sec:cov_buildings}, we introduce practical distance-dependent LOS/NLOS models that are special cases of the general case characterized in Theorem~\ref{th:P_cov} and that will be used in Section~\ref{sec:num} when obtaining numerical results.

\subsection{3GPP LOS/NLOS Model} \label{sec:cov_3gpp}

A widely used distance-dependent LOS/NLOS model is the ITU-R UMi model \cite {UMi} (referred to as 3GPP LOS/NLOS model in the following), which is characterized by the LOS probability function
\begin{equation}
\label{eq:3gpp} p_{\los}(r) = \min \big( \tfrac{18}{r}, 1 \big) \big( 1 - e^{-\tfrac{r}{36}} \big) + e^{-\tfrac{r}{36}}.
\end{equation}
Observe that, using \eqref{eq:3gpp}, the propagation is always in LOS conditions for $r \leq 18$~m. In practice, this implies that for BS densities above $\lambda=10^{-2}$~BSs/m$^{2}$ and closest BS association, the probability of LOS coverage is very close to one and, as a consequence, some NLOS terms in \eqref{eq:P_cov}--\eqref{eq:L_I} can be neglected.

Following this line of thought, we propose a simplified model that can be used for analytical calculations and that is based on the LOS probability function
\begin{align}
\label{eq:3gpp_simpl} p_{\los}(r) = \left\{
\begin{array}{ll}
1, & r \in [0, D) \\
0, & r \in [D, \infty)
\end{array}
\right.
\end{align}
with $D$ being the critical distance below which all BSs are in LOS conditions. The system performance resulting from \eqref{eq:3gpp_simpl} with $D=18$~m in terms of coverage probability very accurately approximates that obtained with the original 3GPP LOS probability function \eqref{eq:3gpp}, as shown by the numerical results in Section~\ref{sec:num}. In this scenario, the coverage probabilities for closest and strongest BS association, given in the general form \eqref{eq:P_cov}, simplify to
\begin{align}
\nonumber \hspace{-2mm} \Pcov^{(\rmC)} (\theta) & = 2 \pi \lambda \bigg( \int_{0}^{D} \sum_{k=0}^{m-1} \bigg[ \frac{(-s)^{k}}{k!} \frac{\diff^{k}}{\diff s^{k}} \widetilde{\setL}_{I} (s) \bigg]_{s = m \theta / \ell_{\los}(r,h)} e^{-\pi \lambda r^{2}} r \diff r \\
\label{eq:P_cov_3gpp_C} & \phantom{=} \ + \int_{D}^{\infty} \setL_{I}^{\nlos} \bigg( \frac{\theta}{\ell_{\nlos}(r,h)} \bigg) e^{-\pi \lambda r^{2}} r \diff r \bigg) \\
\label{eq:P_cov_3gpp_S} \hspace{-2mm} \Pcov^{(\rmS)} (\theta) & = 2 \pi \lambda \bigg( \int_{0}^{D} \sum_{k=0}^{m-1} \bigg[ \frac{(-s)^{k}}{k!} \frac{\diff^{k}}{\diff s^{k}} \widetilde{\setL}_{I} (s) \bigg]_{s = m \theta / \ell_{\los}(r,h)} r \diff r + \int_{D}^{\infty} \widetilde{\setL}_{I} \bigg( \frac{\theta}{\ell_{\nlos}(r,h)} \bigg) r \diff r \bigg)
\end{align}
respectively, with $\setL_{I}^{\nlos}(s)$ defined in \eqref{eq:L_I_nlos} and
\begin{align}
\label{eq:L_I_3gpp} \widetilde{\setL}_{I} (s) & \triangleq \setL_{I}^{\nlos} (s) \exp \bigg( - 2 \pi \lambda \int_{\nu(r)}^{\infty} \bigg( \frac{1}{1 + s \ell_{\nlos}(t,h)} - \frac{1}{(1 + \frac{s}{m} \ell_{\los}(t,h))^{m}} \bigg) t \diff t \bigg).
\end{align}
The coverage probabilities \eqref{eq:P_cov_3gpp_C}--\eqref{eq:P_cov_3gpp_S} can now be evaluated via numerical integration and differentiation, although the latter can be cumbersome in practice, especially for large values of $m$. Thus, to make numerical evaluation more efficient, one can use Proposition~\ref{pro:Alzer} to obtain tractable upper bounds with no derivatives.

\subsection{LOS/NLOS Model with Randomly Placed Buildings} \label{sec:cov_buildings}

In this section we propose a practical model for $p_{\los}(r)$ that takes into account the combined influence of the link distance and the BS height through the probability of the link being blocked by a building. Other options exist in the literature: for instance, in \cite{And16a}, the BS height, the link distance, and the pathloss exponent are related through the effect of the ground-reflected ray.

\begin{figure}[t!]
\centering
\includegraphics[scale=1]{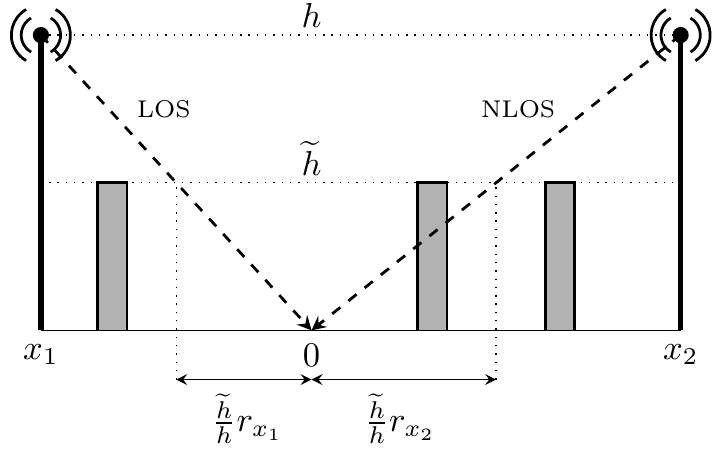}
\caption{LOS/NLOS model with buildings randomly placed between the BSs and the typical UE.} \label{fig:buildings}
\end{figure}

Given a BS located at $x$, we assume that buildings with fixed height $\widetilde{h}$, measured in [m], are randomly placed between $x$ and the typical UE. If the straight line between the elevated BS at $x$ and the typical UE does not cross any buildings, then the transmission occurs in LOS propagation conditions; alternatively, if at least one building cuts this straight line, then the transmission occurs in NLOS propagation conditions.\footnote{The cumulative effect of multiple obstacles is considered in \cite{Lee16}.} A simplified example is illustrated in Figure~\ref{fig:buildings}. Note that, in this context, the probability of $x$ being in LOS propagation conditions depends not only on the distance $r_{x}$, but also on the parameter $\tau \triangleq \min \big( \frac{\widetilde{h}}{h}, 1 \big)$. More precisely, the LOS probability corresponds to the probability of having no buildings in the segment of length $\tau r_{x}$ next to the typical UE.

If the location distribution of the buildings follows a one-dimensional PPP with density $\widetilde{\lambda}$, measured in [buildings/m], the LOS probability function is given by $p_{\los}(r_{x}, \tau) = e^{-\widetilde{\lambda} \tau r_{x}}$. Observe that $p_{\los}(r_{x}, \tau) = 1$ (all links are in LOS propagation conditions) when $\widetilde{\lambda} = 0$ or $\widetilde{h} = 0$, whereas $p_{\los}(r_{x}, \tau) = 0$ (all links are in NLOS propagation conditions) when $\widetilde{\lambda} \to \infty$. In Section~\ref{sec:num}, we will numerically illustrate the effect of different building densities and comment on the interplay between LOS/NLOS desired signal and interference.

\section{The Effect of LOS Fading} \label{sec:los}

In this section, we consider the effect of LOS propagation alone. In doing so, we fix $p_{\los}(r)=1$, $\forall r \in [0, \infty)$ so that all signals from both serving and interfering BSs are subject to Nakagami-$m$ fading. Furthermore, we consider a single pathloss exponent $\alpha$ and we neglect the BS height by fixing $h=0$: under this setting, we have $\ell_{\los} (r,h) = \ell_{\nlos} (r,h) = r^{-\alpha}$. Hence, the results derived in this section implicitly assume non-elevated BSs; in turn, we make the dependence on the shape parameter $m$ explicit in the resulting expressions of the Laplace transform of the interference and coverage probability.

Let us introduce the following preliminary definitions:
\begin{align}
\label{eq:eta} \eta (s,m,r) & \triangleq {}_{2}F_{1} \big( m, - \tfrac{2}{\alpha}, 1 - \tfrac{2}{\alpha}, -\tfrac{s}{m r^{\alpha}} \big) \\
\label{eq:zeta} \zeta (m) & \triangleq - \frac{\Gamma \big( m + \frac{2}{\alpha} \big) \Gamma \big( - \frac{2}{\alpha} \big)}{\alpha \Gamma(m)}
\end{align}
with ${}_{2}F_{1}(a,b,c,z)$ denoting the Gauss hypergeometric function. In addition, we introduce the notation $(z)_{k} \triangleq \frac{\Gamma(z+k)}{\Gamma(z)} = z (z+1) \ldots (z+k-1)$.

\begin{proposition} \rm{
For LOS propagation conditions, the Laplace transforms of the interference for closest and strongest BS association are given by
\begin{align}
\label{eq:L_I_los_C} \setL_{I}^{\los, (\rmC)} (s,m) & \triangleq \exp \big( - \pi \lambda r^{2} \big( \eta(s, m, r) - 1 \big) \big) \\
\label{eq:L_I_los_S} \setL_{I}^{\los, (\rmS)} (s,m) & \triangleq \exp \big( - 2 \pi \lambda \zeta(m) \big( \tfrac{s}{m} \big)^{\frac{2}{\alpha}} \big)
\end{align}
respectively, with $\eta (s, m, r)$ and $\zeta (m)$ defined in \eqref{eq:eta}--\eqref{eq:zeta}. For NLOS propagation conditions, the Laplace transforms of the interference for closest and strongest BS association are given by
\begin{align}
\label{eq:L_I_nlos_C} \setL_{I}^{\nlos, (\rmC)} (s) & \triangleq \setL_{I}^{\los, (\rmC)} (s, 1) \\
\label{eq:L_I_nlos_S} \setL_{I}^{\nlos, (\rmS)} (s) & \triangleq \setL_{I}^{\los, (\rmS)} (s, 1)
\end{align}
respectively, and the corresponding coverage probabilities can be written as
\begin{align}
\label{eq:P_cov_nlos_C} \Pcov^{\nlos, (\rmC)} (\theta) & \triangleq \frac{1}{\eta(\theta,1,1)} \\
\label{eq:P_cov_nlos_S} \Pcov^{\nlos, (\rmS)} (\theta) & \triangleq \frac{1}{2 \zeta(1) \theta^{\frac{2}{\alpha}}}
\end{align} 
respectively.}
\end{proposition}

\begin{IEEEproof}
The Laplace transforms of the interference subject to Nakagami-$m$ fading are obtained by solving \eqref{eq:L_I_nlos} for closest and strongest BS association. Furthermore, the coverage probabilities  in presence of Rayleigh fading can be derived from \cite{And11,Dhi12}, respectively.
\end{IEEEproof} \vspace{1mm}

Let $B_{k} \big( z_{1}, z_{2}, \ldots, z_{k} \big) = \sum_{j=1}^{k} B_{k,j} \big( z_{1}, z_{2}, \ldots, z_{k-j+1} \big)$ denote the $k$th complete Bell polynomial, where $B_{k,j} \big( z_{1}, z_{2}, \ldots, z_{k-j+1} \big)$ is the incomplete Bell polynomial. The following theorem provides closed-form expressions of the coverage probabilities in presence of LOS propagation, i.e., with Nakagami-$m$ fading.

\begin{theorem} \label{th:P_cov_los} \rm{
For LOS propagation conditions, the coverage probability is given as follows.
\begin{itemize}
\item[\textit{(a)}] For closest BS association, we have
\begin{align}
\nonumber \Pcov^{\los, (\rmC)} (\theta,m) & \triangleq \frac{1}{\eta (m \theta, m, 1)} \\
\label{eq:P_cov_los_C} & \hspace{-5mm} \times \bigg( 1 + \sum_{k=1}^{m-1} \sum_{j=1}^{k} \frac{j!}{k!} B_{k,j} \bigg( \frac{\psi_{1} (\theta,m)}{\eta (m \theta, m, 1)}, \frac{\psi_{2} (\theta,m)}{\eta (m \theta, m, 1)}, \ldots, \frac{\psi_{k-j+1} (\theta,m)}{\eta (m \theta, m, 1)} \bigg) \bigg)
\end{align}
with $\eta (s, m, r)$ defined in \eqref{eq:eta} and
\begin{align}
\label{eq:psi} \psi_{k} (\theta, m) \triangleq - \big( - \tfrac{2}{\alpha} \big)_{k} \bigg( \eta (m \theta, m, 1) - \sum_{j=1}^{k} \frac{(m)_{k-j}}{\big( 1 - \tfrac{2}{\alpha}\big)_{k-j}} \theta^{k-j} (1 + \theta)^{-m-k+j} \bigg).
\end{align}
\item[\textit{(b)}] For strongest BS association, we have
\begin{align}
\label{eq:P_cov_los_S} \Pcov^{\los, (\rmS)} (\theta,m) = \Pcov^{\nlos, (\rmS)} (\theta)
\end{align}
with $\Pcov^{\nlos, (\rmS)} (\theta)$ defined in \eqref{eq:P_cov_nlos_S}.
\end{itemize}}
\end{theorem}

\begin{IEEEproof}
See Appendix.~\ref{sec:A1_P_cov_los1}.
\end{IEEEproof} \vspace{1mm}

\begin{remark} \rm{
For strongest BS association, Theorem~\ref{th:P_cov_los}--\textit{(b)} states that Nakagami-$m$ fading does not affect the coverage probability with respect to Rayleigh fading: this stems from the fact that, under LOS fading, the desired signal power grows with the shape parameter $m$ at the same rate as the interference power.}
\end{remark}

\noindent The expression in Theorem~\ref{th:P_cov_los}--\textit{(a)}, although in closed form, is quite involved: in this respect, Corollary~\ref{cor:P_cov_los} formally characterizes the trend of the coverage probability for LOS propagation conditions and closest BS association.

\begin{corollary} \label{cor:P_cov_los} \rm{
For LOS propagation conditions and closest BS association, recalling the definition of $\Pcov^{\nlos, (\rmS)} (\theta)$ in \eqref{eq:P_cov_nlos_S}, the following hold:
\begin{itemize}
\item[\textit{(a)}] $\Pcov^{\los, (\rmC)} (\theta,m+1) > \Pcov^{\los, (\rmC)} (\theta,m) > \Pcov^{\nlos, (\rmC)} (\theta)$, $\forall m \geq 1$;
\item[\textit{(b)}] $\lim_{m \to \infty} \Pcov^{\los, (\rmC)} (\theta,m) = \Pcov^{\nlos, (\rmS)} (\theta)$.
\end{itemize}}
\end{corollary}

\begin{IEEEproof}
See Appendix.~\ref{sec:A1_P_cov_los2}.
\end{IEEEproof} \vspace{1mm}

\begin{remark} \rm{
For closest BS association, Corollary~\ref{cor:P_cov_los} highlights the beneficial effect of Nakagami-$m$ fading on the coverage probability: this stems from the fact that, under LOS fading, the desired signal power grows at a higher rate than the interference power. In addition, as the shape parameter $m$ increases, the performance with closest BS association converges that with strongest BS association.}
\end{remark}

\section{The Effect of BS Height} \label{sec:height}

We now focus on the effect of BS height on the coverage probability. In doing so, we set the shape parameter $m=1$ and, as in Section~\ref{sec:los}, we consider a single pathloss exponent $\alpha$, which yields $\ell_{\los} (r,h) = \ell_{\nlos} (r,h) = (r+h)^{-\frac{\alpha}{2}}$. Hence, the results derived in this section implicitly assume that all signals from both serving and interfering BSs are subject to Rayleigh fading. While the coverage probability for both closest and strongest BS association is independent on the BS density $\lambda$ when $h=0$, as can be observed from \eqref{eq:P_cov_nlos_C}--\eqref{eq:P_cov_nlos_S} (see also \cite{And11} and \cite{Dhi12}, respectively, for details), the impact of BS height becomes visible as $\lambda$ increases \cite{Din16a}. Therefore, we make the dependence on $\lambda$ explicit in the resulting expressions of the coverage probability and the ASE.

\subsection{Impact on Interference} \label{sec:height_int}

The interference with elevated BSs is characterized here. We begin by observing that $\ell (r_{x}, h)$ yields a bounded pathloss model for any BS height $h > 0$, since BS $x$ cannot get closer than $h$ to the typical UE (this occurs when $r_{x} = 0$). The following lemma expresses the Laplace transforms of the interference for a fixed BS height $h$.
\begin{lemma} \label{lem:L_I_0} {\rm
For elevated BSs, the Laplace transforms of the interference for closest and strongest BS association can be written as
\begin{align}
\label{eq:L_I_0_C} \setL_{I}^{(\rmC)} (s) & \triangleq \setL_{I}^{\nlos, (\rmC)}(s) \exp \bigg( 2 \pi \lambda \int_{r}^{\sqrt{r^{2}+h^{2}}} \bigg( 1 - \frac{1}{1 + s t^{-\alpha}} \bigg) t \diff t \bigg) \\
& = \exp \big( - \pi \lambda (r^{2} + h^{2}) \big( \eta \big( s, 1, \sqrt{r^{2} + h^{2}} \big) - 1 \big) \big) \\
\label{eq:L_I_0_S} \setL_{I}^{(\rmS)} (s) & \triangleq \setL_{I}^{\nlos, (\rmS)}(s) \exp \bigg( 2 \pi \lambda \int_{0}^{h} \bigg( 1 - \frac{1}{1 + s t^{-\alpha}} \bigg) t \diff t \bigg) \\
& = \exp \big( - \pi \lambda h^{2} \big( \eta(s, 1, h) - 1 \big) \big)
\end{align}
respectively, where $\setL_{I}^{\nlos, (\rmC)}(s)$ and $\setL_{I}^{\nlos, (\rmS)}(s)$ are the Laplace transforms of the interference with non-elevated BSs defined in \eqref{eq:L_I_nlos_C}--\eqref{eq:L_I_nlos_S}.}
\end{lemma}

\begin{IEEEproof}
The Laplace transforms of the interference \eqref{eq:L_I_0_C} and \eqref{eq:L_I_0_S} can be obtained from \eqref{eq:L_I_nlos} first by substituting $\sqrt{t^{2} + h^{2}} \to t_{h}$ and then by splitting the integration intervals in two parts; see also Appendix~\ref{sec:A2_height}.
\end{IEEEproof} \vspace{1mm}

\begin{remark} \rm{
From \eqref{eq:L_I_0_C} and \eqref{eq:L_I_0_S}, it is straightforward to see that the interference is reduced when $h>0$ with respect to when $h=0$, since the original Laplace transforms are multiplied by exponential terms with positive arguments.}
\end{remark}

For strongest BS, we provide a further interesting result on the expected interference power. Recall that, for strongest BS association and for $h=0$, the expected interference power is infinite \cite[Ch.~5.1]{Hae12}. Let $U(a,b,z) \triangleq \frac{1}{\Gamma(a)} \int_{0}^{\infty} e^{-z t} t^{a - 1} (1 + t)^{b - a - 1} \diff t$ denote Tricomi's confluent hypergeometric function and let $E_{n}(z) \triangleq \int_{1}^{\infty} e^{-z t} t^{-n} \diff t$ be the exponential integral function. A consequence of the bounded pathloss model is given in the following lemma, which characterizes the expected interference with elevated BSs for strongest BS association.

\begin{lemma} \label{lem:int} \rm{
For elevated BSs, the expected interference power for strongest BS association is finite and is given by
\begin{align} \label{eq:int}
\Exp \bigg[ \sum_{y \in \Phi \backslash \{x\}} g_{y} \ell(r_{y},h) \bigg] < \sum_{i=1}^{\infty} (\pi \lambda)^{i} h^{2 i - \alpha} U \big( i, i + 1 - \tfrac{\alpha}{2}, \pi \lambda h^{2} \big)
\end{align}
where the expected interference power from the nearest interfering BS, whose location is denoted by $x_{1}$, corresponds to
\begin{align} \label{eq:int_nearest}
\Exp \big[ g_{x_{1}} \ell(r_{y_{1}},h) \big] = \pi \lambda h^{2 - \alpha} e^{\pi \lambda h^{2}} E_{\frac{\alpha}{2}}(\pi \lambda h^{2}).
\end{align}}
\end{lemma}

\begin{IEEEproof}
See Appendix~\ref{sec:A2_int}.
\end{IEEEproof} \vspace{1mm}

\subsection{Impact on Coverage Probability and ASE} \label{sec:height_cov}

We now focus on the effect of BS height on the coverage probability and on the ASE. Theorem~\ref{th:height} provides the coverage probabilities for a fixed BS height $h$.

\begin{theorem} \label{th:height} \rm{
For elevated BSs, recalling the definition of $\eta(s, m, r)$ in \eqref{eq:eta}, the coverage probability is given as follows.
\begin{itemize}
\item[\textit{(a)}] For closest BS association, we have
\begin{align}
\label{eq:P_cov_C} \Pcov^{(\rmC)} (\theta, \lambda) & \triangleq \Pcov^{\nlos,(\rmC)} (\theta) \exp \big( - \pi \lambda h^{2} \big( \eta(\theta, 1, 1) - 1 \big) \big)
\end{align}
where $\Pcov^{\nlos,(\rmC)} (\theta)$ is the coverage probability with non-elevated BSs defined in \eqref{eq:P_cov_nlos_C}.
\item[\textit{(b)}] For strongest BS association, we have
\begin{align}
\label{eq:P_cov_S} \Pcov^{(\rmS)} (\theta, \lambda) & \triangleq 2 \pi \lambda \int_{h}^{\infty} \exp \big( - \pi \lambda h^{2} \big( \eta(\theta r^{\alpha}, 1, h) - 1 \big) \big) r \diff r.
\end{align}
\end{itemize}}
\end{theorem}

\begin{IEEEproof}
See Appendix~\ref{sec:A2_height}.
\end{IEEEproof} \vspace{1mm}

\noindent Notably, for closest BS association, a closed-form expression is available; therefore, Corollary~\ref{cor:lambda_opt} gives the optimal BS density in terms of achievable ASE.

\begin{corollary} \label{cor:lambda_opt} \rm{
For elevated BS and closest BS association, let $\ASE^{(\rmC)}(\theta, \lambda) \triangleq \lambda \Pcov^{(\rmC)} (\theta, \lambda) \log_{2} (1 + \theta)$ denote the achievable ASE. Then, the optimal BS density is given by
\begin{align} \label{eq:lambda_opt}
\lambda_{\mathrm{opt}}^{(\rmC)} \triangleq \argmax_{\lambda} \ASE^{(\rmC)}(\theta, \lambda) = \frac{1}{\pi h^{2} \big( \eta(\theta, 1, 1) - 1 \big)}
\end{align}
and the maximum ASE corresponds to
\begin{align}
\ASE_{\mathrm{max}}^{(\rmC)}(\theta) \triangleq \ASE^{(\rmC)}(\theta, \lambda_{\mathrm{opt}}^{(\rmC)}) = \frac{e^{-1}}{\pi h^{2} \eta(\theta, 1, 1) \big( \eta(\theta, 1, 1) - 1 \big)}.
\end{align}}
\end{corollary}

\begin{IEEEproof}
The optimal BS density $\lambda_{\mathrm{opt}}^{(\rmC)}$ is simply obtained as the solution of $\frac{\diff}{\diff \lambda} \lambda \Pcov^{(\rmC)} (\theta, \lambda) = 0$.
\end{IEEEproof} \vspace{1mm}

Theorem~\ref{th:height} unveils the detrimental effect of BS height on the system performance. This degradation stems from the fact that the distance of the typical UE from its serving BS is more affected by the BS height than the distances from the interfering BSs and, therefore, desired signal power and interference power do not grow at the same rate as in the case with $h = 0$. The following corollary strengthens this claim by showing the asymptotic performance for both closest and strongest BS association.

\begin{corollary} \label{cor:height_limits} \rm{
For elevated BSs, recalling the definitions of the coverage probabilities with non-elevated BSs $\Pcov^{\nlos,(\rmC)} (\theta)$ and $\Pcov^{\nlos,(\rmS)} (\theta)$ in \eqref{eq:P_cov_nlos_C}--\eqref{eq:P_cov_nlos_S}, the following holds:
\begin{itemize}
\item[\textit{(a)}] $\displaystyle \lim_{\lambda \to 0} \Pcov^{(\rmC)} (\theta, \lambda) = \Pcov^{\nlos,(\rmC)} (\theta)$;
\item[\textit{(b)}] $\displaystyle \lim_{\lambda \to 0} \Pcov^{(\rmS)} (\theta, \lambda) = \Pcov^{\nlos,(\rmS)} (\theta)$;
\item[\textit{(c)}] $\displaystyle \lim_{\lambda \to \infty} \Pcov^{(\rmC)} (\theta, \lambda) = \lim_{\lambda \to \infty} \Pcov^{(\rmS)} (\theta,\lambda) = 0$.
\end{itemize}
}
\end{corollary}

\begin{IEEEproof}
See Appendix~\ref{sec:A2_limits}.
\end{IEEEproof} \vspace{1mm}

\begin{remark} \rm{
For a fixed BS height $h > 0$, the coverage probability monotonically decreases as the BS density $\lambda$ increases, eventually leading to near-universal outage: as a consequence, the ASE also decays to zero as $\lambda \to \infty$. On the other hand, the effect of BS height becomes negligible as $\lambda \to 0$.}
\end{remark}

\noindent In practice, we will see in Section~\ref{sec:num} that the coverage probability and the ASE decay to zero even for moderately low BS densities (i.e., for $\lambda \simeq 10^{-2}$~BSs/m$^2$).

\begin{remark} \label{rem:lambda_opt} \rm{
For a fixed BS density $\lambda$, the coverage probability monotonically decreases as the BS height $h$ increases. More specifically, from Corollary~\ref{cor:lambda_opt}, we have that $\ASE_{\mathrm{max}}^{(\rmC)}(\theta) \propto \frac{1}{h^{2}}$. Therefore, the optimal BS height is $h=0$.}
\end{remark}

\noindent When serving and interfering BSs are characterized by the same propagation conditions or, more generally, by the same distance-dependent LOS probability function (as the one described in Section~\ref{sec:cov_buildings}), the optimal BS height is always $h = 0$, which confirms the findings in \cite{Din16a}. However, under a propagation model where the interfering BSs are always in NLOS propagation conditions and the serving BS can be either in LOS or NLOS propagation conditions, a non-zero optimal BS height is expected: in fact, in this case, there would be a tradeoff between pathloss (for which a low BS is desirable) and probability of LOS desired signal (for which a high BS is desirable).

\section{Numerical Results and Discussion} \label{sec:num}

In this section, we present numerical results to assess our theoretical findings. In particular,
we aim at answering the following general question: \textit{what is the individual and combined effect of LOS/NLOS fading, LOS/NLOS pathloss, and BS height on the UE and network performance?}

\subsection{The Effect of LOS Fading} \label{sec:num_1}

\begin{figure}[t!]
\centering
\includegraphics[scale=0.9]{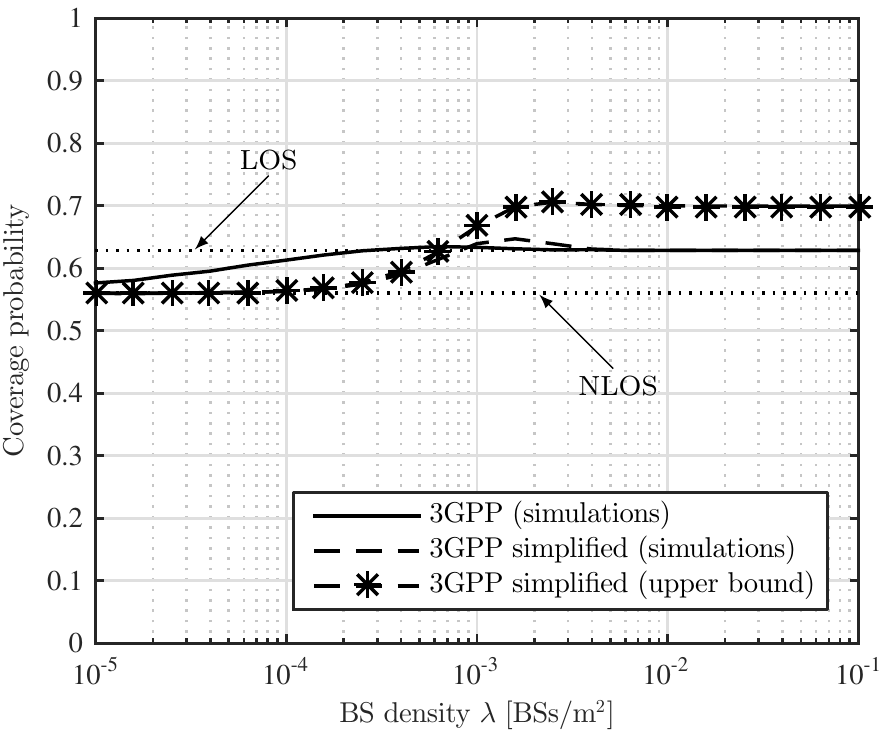}
\includegraphics[scale=0.9]{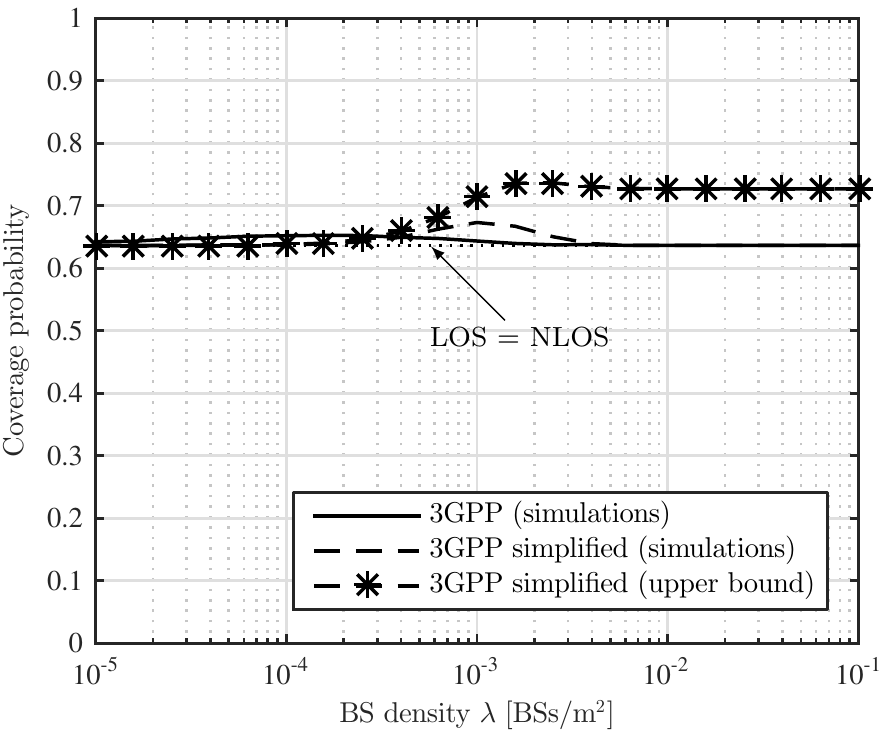}
\caption{Coverage probability with the 3GPP LOS/NLOS model (cf. Section~\ref{sec:cov_3gpp}), $m=10$, $\alpha=4$, and non-elevated BSs against BS density $\lambda$ for closest (left) and strongest (right) BS association.} \label{fig:3gpp} \vspace{-3mm}
\end{figure}

We begin by examining the distance-dependent 3GPP LOS/NLOS model presented in Section~\ref{sec:cov_3gpp}. Considering a shape parameter $m=10$ (which corresponds to a Rician $K$-factor $K \simeq 13$~dB), a single pathloss exponent $\alpha = 4$, non-elevated BSs ($h=0$), and SIR threshold $\theta = 0$~dB, Figure~\ref{fig:3gpp} illustrates the coverage probability based on the LOS probability function \eqref{eq:3gpp} for closest and strongest BS association against the BS density $\lambda$. The coverage probability based on the simplified LOS probability function \eqref{eq:3gpp_simpl} with $D=18$~m and the corresponding upper bound, obtained by applying Proposition~\ref{pro:Alzer} followed by numerical integration, are also plotted. In accordance with Corollary~\ref{cor:3gpp}, the coverage probability corresponds to the NLOS case for low BS densities (i.e., $\lambda \leq 10^{-4}$~BSs/m$^{2}$) and to the LOS case for high BS densities (i.e., $\lambda \geq 10^{-2}$~BSs/m$^{2}$): in particular, these two cases coincide for strongest BS association, as stated in Theorem~\ref{th:P_cov_los}--\textit{(b)}. Furthermore, the upper bound is remarkably tight for low BS densities and, in general, tighter for closer BS association than for strongest BS association.

\begin{figure}[t!]
\centering
\includegraphics[scale=0.9]{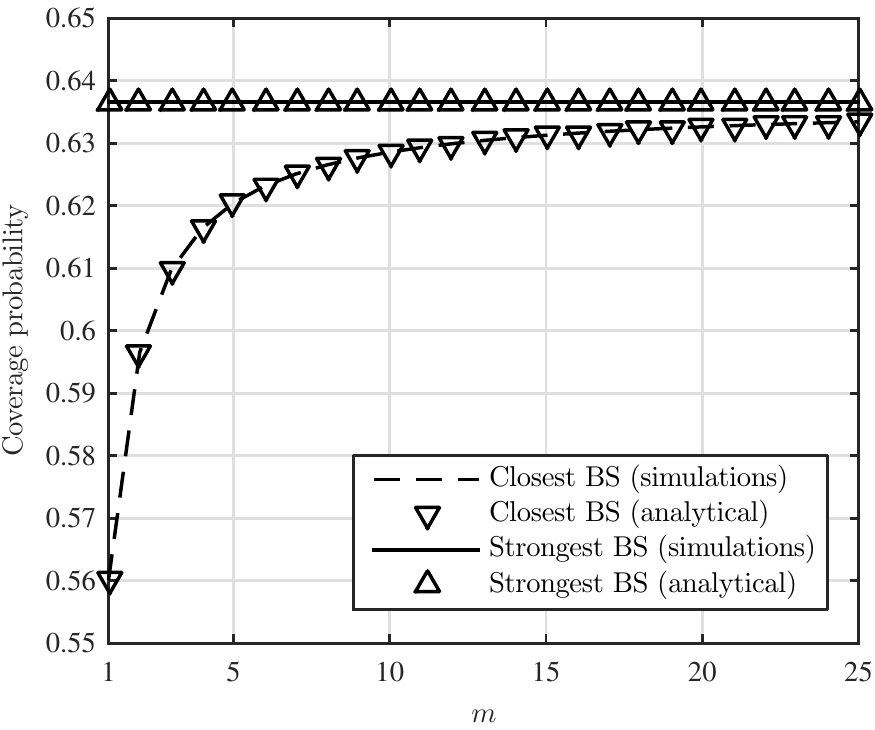}
\caption{Coverage probability with Nakagami-$m$ fading (cf. Section~\ref{sec:los}), $\alpha=4$, and non-elevated BSs against shape parameter $m$.} \label{fig:LOS} \vspace{-3mm}
\end{figure}

Now, let us consider the LOS setting of Section~\ref{sec:los}, i.e., all signals from both serving and interfering BSs are subject to Nakagami-$m$ fading, with a single pathloss exponent $\alpha = 4$, non-elevated BSs ($h=0$), and SIR threshold $\theta = 0$~dB. Figure~\ref{fig:LOS} plots the coverage probability for closest and strongest BS association against the shape parameter of the Nakagami-$m$ fading. Firstly, the analytical expressions derived in Theorem~\ref{th:P_cov_los} match the numerical curves exactly. Secondly, in accordance with Corollary~\ref{cor:P_cov_los}, the coverage probability for closest BS association increases with $m$ until it approaches the (constant) value of the coverage probability with strongest BS association: in particular, the two values are approximately the same already for $m=25$.

\subsection{The Effect of BS Height} \label{sec:num_2}

\begin{figure}[t!]
\begin{minipage}[c]{0.49\textwidth}
\centering
\vspace{1.8mm}
\includegraphics[scale=0.9]{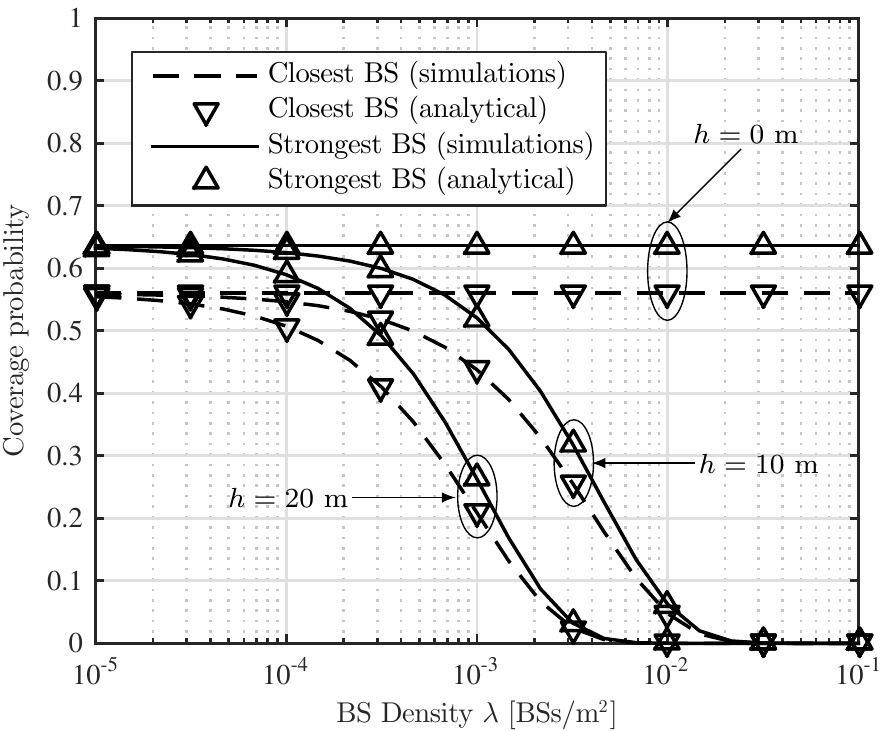}
\caption{Coverage probability with elevated BSs (cf. Section~\ref{sec:height}), $m=1$, and $\alpha=4$ against BS density $\lambda$.} \label{fig:height_Pcov}
\end{minipage}
\hspace{1mm}
\begin{minipage}[c]{0.49\textwidth}
\centering
\includegraphics[scale=0.9]{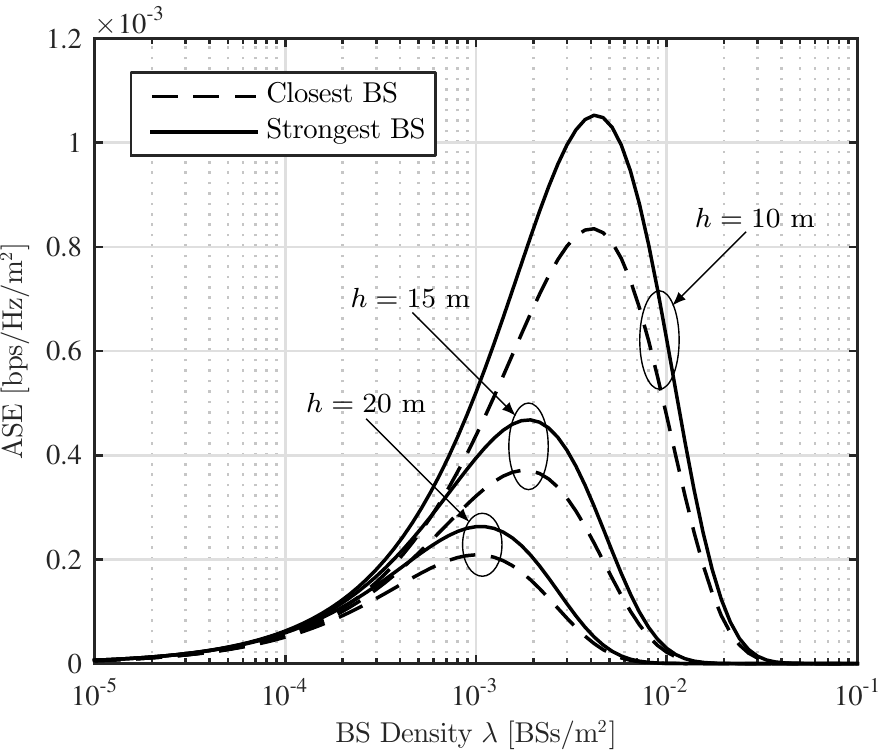}
\caption{Achievable ASE with elevated BSs (cf. Section~\ref{sec:height}), $m=1$, and $\alpha=4$ against BS density $\lambda$.} \label{fig:height_ASE}
\end{minipage}
\end{figure}

Here, we focus on the effect of BS height alone and, as in Section~\ref{sec:height}, we set the shape parameter $m=1$ and consider a single pathloss exponent $\alpha = 4$ and SIR threshold $\theta = 0$~dB. Figure~\ref{fig:height_Pcov} plots the coverage probability with elevated BSs for closest and strongest BS association against the BS density $\lambda$; two BS heights are considered, i.e., $h=10$~m and $h=20$~m. First of all, the analytical expressions derived in Theorem~\ref{th:height} match the numerical curves exactly. Interestingly, it is shown that the coverage probability decays to zero even for moderately low BS densities, i.e., at $\lambda \simeq 10^{-2}$~BSs/m$^{2}$ with $h=20$~m and at $\lambda \simeq 3 \times 10^{-2}$~BSs/m$^{2}$ with $h=10$~m. For comparison, the coverage probabilities with $h=0$~m, i.e., $\Pcov^{\nlos,(\rmC)} (\theta)$ and $\Pcov^{\nlos,(\rmS)} (\theta)$ in \eqref{eq:P_cov_nlos_C}--\eqref{eq:P_cov_nlos_S}, are also depicted. In accordance with Corollary~\ref{cor:height_limits}, the coverage probability with elevated BSs converges to that with $h=0$ as $\lambda \to 0$: this is already verified at $\lambda \simeq 10^{-5}$~BSs/m$^{2}$, when the BSs are so far away from the typical UE that the effect of BS height become negligible.

In Figure~\ref{fig:height_ASE}, we show the achievable ASE with elevated BSs for closest and strongest BS association against the BS density $\lambda$; three BS heights are considered, i.e., $h=10$~m, $h=15$~m, and $h=20$~m. Here, the detrimental effect of BS height on the system performance appears even more evident. As an example, considering closest BS association, the maximum ASE is $0.84 \times 10^{-3}$~bps/Hz/m$^{2}$ for $h=10$~m and $0.21 \times 10^{-3}$~bps/Hz/m$^{2}$ for $h=20$~m; likewise, considering strongest BS association, the maximum ASE is $1.04 \times  10^{-3}$~bps/Hz/m$^{2}$ for $h=10$~m and $0.26 \times 10^{-3}$~bps/Hz/m$^{2}$ for $h=20$~m. Hence, in accordance with Corollary~\ref{cor:lambda_opt}, doubling the BS height reduces the maximum ASE by a factor of four (see also Remark~\ref{rem:lambda_opt}). Furthermore, it is worth noting that the BS density that maximizes the ASE for closest BS association, (i.e., $\lambda_{\mathrm{opt}}^{(\rmC)}$, which is derived in closed form in \eqref{eq:lambda_opt}) coincides with the optimal BS density for the case of strongest BS association: this corresponds to $\lambda \simeq 4 \times 10^{-3}$~BSs/m$^{2}$ for $h=10$~m and to $\lambda \simeq 10^{-3}$~BSs/m$^{2}$ for $h=20$~m.

\subsection{The General Case} \label{sec:num_3}

\begin{figure}[t!]
\centering
\hspace{-3mm} \includegraphics[scale=0.93]{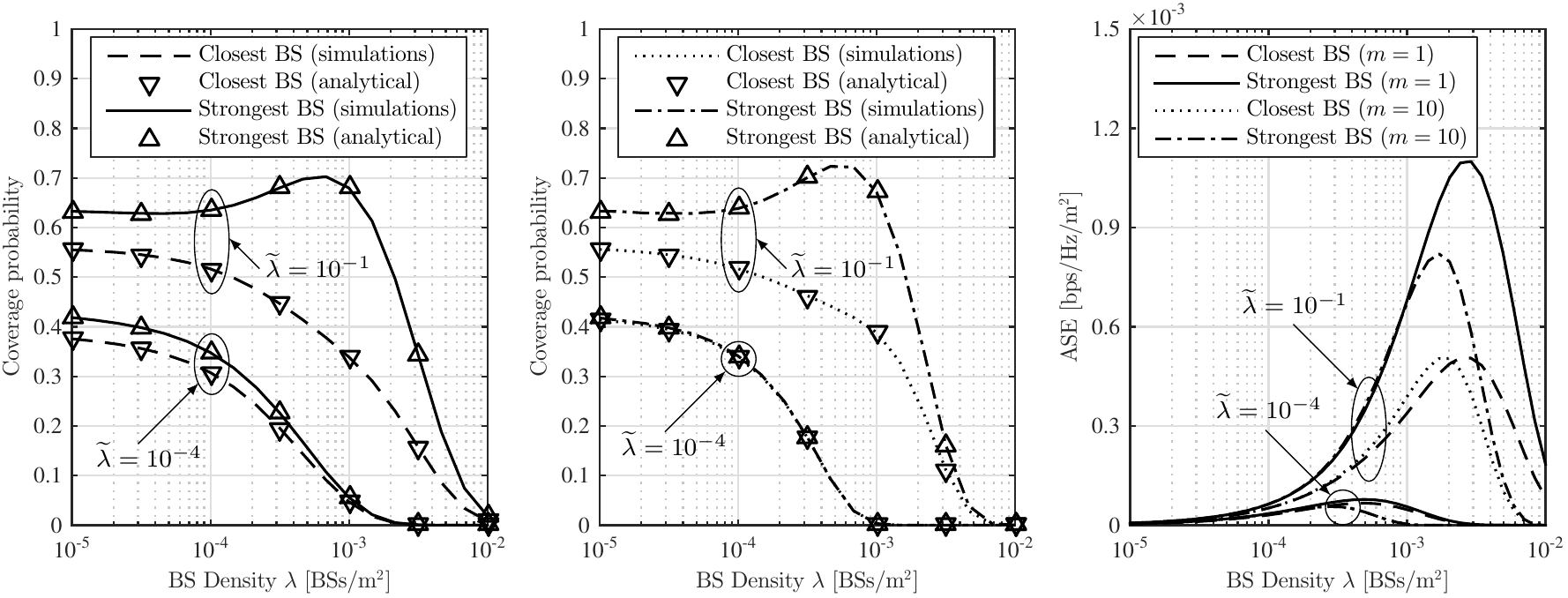}
\caption{Coverage probability with the LOS/NLOS model with randomly placed buildings (c.f. Section~\ref{sec:cov_buildings}), $\alpha_{\los}=3$, $\alpha_{\nlos}=4$, $h=20$~m, and $\widetilde{h}=10$~m against BS density $\lambda$ for $m=1$ (left) and $m=10$ (center); corresponding achievable ASE (right).} \label{fig:fig_B}
\end{figure}

Lastly, we use the distance-dependent LOS/NLOS model with randomly placed buildings presented in Section~\ref{sec:cov_buildings} to analyze the combined effect of LOS/NLOS fading, LOS/NLOS pathloss, and BS height. Considering pathloss exponents $\alpha_{\los}=3$ and $\alpha_{\nlos}=4$, elevated BSs with $h=20$~m, building height $\widetilde{h}=10$~m, building densities $\widetilde{\lambda} = 10^{-4}$~buildings/m and $\widetilde{\lambda} = 10^{-1}$~buildings/m, and SIR threshold $\theta = 0$~dB, Figure~\ref{fig:fig_B} illustrates the coverage probability and the achievable ASE for closest and strongest BS association against the BS density $\lambda$; two shape parameters of the Nakagami-$m$ fading are considered, i.e., $m=1$ and $m=10$. Observing the curves in Figure~\ref{fig:fig_B}, the detrimental effects on the system performance can be ranked in decreasing order of importance as follows:
\begin{itemize}
\item[\textit{1)}] The BS height is evidently the dominant effect, since it eventually leads to near-universal outage regardless of the other parameters and even for moderately low BS densities (cf. Figures~\ref{fig:height_Pcov}--\ref{fig:height_ASE}).
\item[\textit{2)}] The LOS pathloss, which is determined by the building density, is the second most important effect: specifically, a high building density is beneficial in this setting since it creates nearly NLOS pathloss conditions, whereas a low building density leads to a nearly LOS scenario.
\item[\textit{3)}] The LOS fading, also determined by the building density, has a minor impact as compared to the pathloss: in particular, for closest BS association, the coverage with $m=10$ is slightly improved at low BS densities and marginally deteriorated at high BS densities with respect to the case with $m=1$.
\end{itemize}
Moreover, for $\widetilde{\lambda} = 10^{-1}$~buildings/m and strongest BS association, we observe a peak in the coverage probability around $\lambda = 10^{-3}$~BSs/m$^{2}$ due to the choice of the parameters $h$ and $\widetilde{h}$.

\section{Conclusions} \label{sec:concl}

In this paper, we study the downlink performance of dense cellular networks with elevated BSs and LOS/NLOS small-scale and large-scale fading. We introduce a stochastic geometry based framework that accommodates both closest and strongest BS association, dual-slope pathloss, and Nakagami-$m$ fading for the LOS small-scale fading. First, we consider two special--yet practically relevant--cases of distance-dependent LOS/NLOS models, i.e., a 3GPP inspired model and a newly proposed model with randomly placed buildings. Second, considering the effect of LOS propagation alone, we derive closed-form expressions of the coverage probability with Nakagami-$m$ fading. Interestingly, the coverage probability for strongest BS association is the same as in the case of Rayleigh fading, whereas it monotonically increases with the shape parameter $m$ for closest BS association. Lastly, we focus on the effect of elevated BSs and show that the maximum ASE is proportional to the inverse of the squared BS height. Therefore, densifying the network leads to near-universal outage regardless of the other parameters and even at moderately low BS densities.

Further extensions to this framework may include incorporating spatial models with repulsion or minimum separation between BSs (such as Ginibre and Matern processes, respectively) and clustered point processes, as well as non-homogeneous UE distributions. Other factors affecting the performance of real-world networks, such as shadow fading, antenna sectorization, and directional antennas can be also considered in future work.


\appendices

\section{Coverage Probability} \label{sec:A1}
\subsection{Proof of Theorem~\ref{th:P_cov}} \label{sec:A1_P_cov}

The coverage probability is given by
\begin{align}
\Pcov (\theta) & = \Pr \bigg[ \frac{g_{x} \ell_{\rmQ}(r_{x},h)}{I} > \theta \bigg] \\
& = \int_{0}^{\infty} \Pr \bigg[ g_{x} > \frac{\theta I}{\ell_{\rmQ}(r,h)} \Big| r \bigg] f_{r_{x}}(r) \diff r \\
\label{eq:P_cov1} & = \int_{0}^{\infty} \bigg( p_{\los}(r) \Exp_{I} \bigg[ \bar{F}_{\los} \bigg( \frac{\theta I}{\ell_{\los}(r,h)} \bigg) \bigg] + \big( 1 - p_{\los}(r) \big) \Exp_{I} \bigg[ \bar{F}_{\nlos} \bigg( \frac{\theta I}{\ell_{\nlos}(r,h)} \bigg) \bigg] \bigg) f_{r_{x}}(r) \diff r
\end{align}
where \eqref{eq:P_cov1} derives from the fact that $\mathrm{Q} = \mathrm{LOS}$ with probability $p_\los(r)$ and $\mathrm{Q} = \mathrm{NLOS}$ with probability $1 - p_\los(r)$; recall that $\bar{F}_{\los}(z)$ is the CCDF of $g_{x}$ for LOS propagation conditions defined in \eqref{eq:ccdf_los}, whereas the CCDF of $g_{x}$ for NLOS propagation conditions $\bar{F}_{\nlos}(z)$ can be obtained from $\bar{F}_{\los}(z)$ by setting $m=1$. Then, the expression in \eqref{eq:P_cov} readily follows from
\begin{align}
\label{eq:E_los} \Exp_{I} & \big[ \bar{F}_{\los}(z I) \big] = \Exp_{I} \bigg[  e^{-m z I} \sum_{k=0}^{m-1} \frac{(m z)^k}{k!} I^{k} \bigg] = \sum_{k=0}^{m - 1} \bigg[ \frac{(-s)^{k}}{k!} \frac{\diff^{k}}{\diff s^{k}} \setL_{I}(s) \bigg]_{s = m z} \\
\label{eq:E_nlos} \Exp_{I} & \big[ \bar{F}_{\nlos}(z I) \big] = \Exp_{I} [e^{-z I}] = \setL_{I}(z).
\end{align}

On the other hand, the Laplace transform in \eqref{eq:L_I} is obtained as
\begin{align}
\setL_{I} (s) & = \Exp_{I} [e^{-s I}] \\
\label{eq:L_I1} & = \Exp_{\Phi} \bigg[ \prod_{y \in \Phi_{\los} \backslash \{x\}} \Exp_{g_{y}} \big[ \exp \big( - s g_{y} \ell_{\los}(r_{y},h) \big) \big] \prod_{y \in \Phi_{\nlos} \backslash \{x\}} \Exp_{g_{y}} \big[ \exp \big( - s g_{y} \ell_{\nlos}(r_{y},h) \big) \big] \bigg] \\
\label{eq:L_I2} & = \Exp_{\Phi} \bigg[ \prod_{y \in \Phi_{\los} \backslash \{x\}} \frac{1}{(1 + \frac{s}{m} \ell_{\los}(r_{y},h))^{m}} \prod_{y \in \Phi_{\nlos} \backslash \{x\}} \frac{1}{1 + s \ell_{\nlos}(r_{y},h)} \bigg] \\
\label{eq:L_I3} & = \exp \bigg( \! - 2 \pi \lambda \int_{\nu(r)}^{\infty} \bigg( 1 - p_{\los}(r) \frac{1}{(1 + \frac{s}{m} \ell_{\los}(t,h))^{m}} - \big( 1 - p_{\los}(r) \big) \frac{1}{1 + s \ell_{\nlos}(t,h)} \bigg) t \diff t \bigg)
\end{align}
where \eqref{eq:L_I2} results from applying the moment generating function of the Gamma and exponential distributions to the first and second expectation terms in \eqref{eq:L_I1}, respectively, and in \eqref{eq:L_I3} we have used the probability generating functional of a PPP. Finally, the expression in \eqref{eq:L_I} is obtained by including \eqref{eq:L_I_nlos} into \eqref{eq:L_I3}, and this completes the proof. \hfill \IEEEQED

\subsection{Multi-Antenna BSs} \label{sec:A1_MISO}

Suppose that the BSs are equipped with $N_{\mathrm{T}}$ transmit antennas and adopt maximum ratio transmission (MRT) beamforming to serve their associated UEs. If the serving BS is located at $x$, the channel power fading gain $g_{x}$ follows the Gamma distribution $\Gamma \big( N_{\mathrm{T}} m, \frac{1}{m} \big)$ if $x \in \Phi_{\los}$ and the chi-squared distribution $\chi_{2 N_{\mathrm{T}}}^{2}$ if $x \in \Phi_{\nlos}$, with CCDFs given by
\begin{align}
\label{eq:ccdf_MISO} \bar{F}_{\los}(z) = e^{-m z} \sum_{k=0}^{N_{\mathrm{T}} m-1} \frac{(m z)^k}{k!}, \qquad \bar{F}_{\nlos}(z) = e^{- z} \sum_{k=0}^{N_{\mathrm{T}}-1} \frac{z^k}{k!}
\end{align}
respectively. On the other hand, the channel power gains from the interfering BSs are distributed as in Section~\ref{sec:SM_ch} or, equivalently as in \eqref{eq:ccdf_MISO} with $N_{\mathrm{T}} = 1$. Then, the coverage probability \eqref{eq:P_cov} becomes
\begin{align}
\nonumber \Pcov (\theta) = \int_{0}^{\infty} \bigg( p_{\los}(r) & \sum_{k=0}^{N_{\mathrm{T}} m-1} \bigg[ \frac{(-s)^{k}}{k!} \frac{\diff^{k}}{\diff s^{k}} \setL_{I} (s) \bigg]_{s = m \theta / \ell_{\los}(r,h)} \\
\label{eq:P_cov_MISO} & + \big( 1 - p_{\los}(r) \big) \sum_{k=0}^{N_{\mathrm{T}}-1} \bigg[ \frac{(-s)^{k}}{k!} \frac{\diff^{k}}{\diff s^{k}} \setL_{I} (s) \bigg]_{s = \theta / \ell_{\nlos}(r,h)} \bigg) f_{r_{x}}(r) \diff r
\end{align}
where $\setL_{I} (s)$ is the Laplace transform of the interference defined in \eqref{eq:L_I}; we refer to Appendix~\ref{sec:A1_P_cov} and \cite[App.~I-B]{Atz17} for details.

\subsection{Proof of Theorem~\ref{th:P_cov_los}} \label{sec:A1_P_cov_los1}

The proof is quite involved. Building on \eqref{eq:P_cov}, the coverage probability in presence of LOS propagation is given by
\begin{align}
\label{eq:P_cov_los} \Pcov^{\los} (\theta,m) & = \sum_{k=0}^{m-1} \int_{0}^{\infty} \bigg[ \frac{(-s)^{k}}{k!} \frac{\diff^{k}}{\diff s^{k}} \setL_{I}^{\los} (s) \bigg]_{s = m \theta r^{\alpha}} f_{r_{x}}(r) \diff r.
\end{align}

\textit{(a)} Let us focus on closest BS association and let us recall the definition of $\setL_{I}^{\los,(\rmC)} (s)$ in \eqref{eq:L_I_los_C}. In addition, let us recall the property of the derivatives of the Gauss hypergeometric function, by which $\frac{\diff^{k}}{\diff z^{k}} {}_{2}F_{1} (a,b,c,z) = \frac{(a)_{k} (b)_{k}}{(c)_{k}} {}_{2}F_{1} (a+k,b+k,c+k,z)$; furthermore, when $c=b+1$, we build on \cite[Eq.~9.137(11)]{Gra07} to obtain $\frac{\diff}{\diff z} {}_{2}F_{1} (a,b,b+1,z) = \frac{b}{z} \big( (1-z)^{-a} - {}_{2}F_{1} (a,b,b+1,z) \big)$, which allows us to write the $k$th derivative of $\eta (s, m, r)$ in \eqref{eq:eta} as
\begin{align}
\frac{\diff^{k}}{\diff s^{k}} \eta (s, m, r) = (-s)^{-k} \big( - \tfrac{2}{\alpha} \big)_{k} \bigg( \eta (s, m, r) - \sum_{j=1}^{k} \frac{(m)_{k-j}}{\big( 1 - \tfrac{2}{\alpha}\big)_{k-j}} \big(\tfrac{s}{m r^{\alpha}} \big)^{k-j} \big( 1 + \tfrac{s}{m r^{\alpha}} \big)^{-m-k+j} \bigg).
\end{align}
For a given $k \geq 1$, we can now derive the $k$th term of the summation in \eqref{eq:P_cov_los} as
\begin{align}
\label{eq:P_cov_nlos_C1} 2 \pi & \lambda \int_{0}^{\infty} \bigg[ \frac{(-s)^{k}}{k!} \frac{\diff^{k}}{\diff s^{k}} \setL_{I}^{\los,(\rmC)} (s) \bigg]_{s = m \theta r^{\alpha}} e^{-\pi \lambda r^{2}} r \diff r \\
& = 2 \pi \lambda \int_{0}^{\infty} \bigg[ \frac{(-s)^{k}}{k!} \frac{\diff^{k}}{\diff s^{k}} \exp \big( - \pi \lambda r^{2} \eta (s,m,r) \big) \bigg]_{s = m \theta r^{\alpha}} r \diff r \\
\nonumber & = 2 \pi \lambda \int_{0}^{\infty} \bigg[ \frac{(-s)^{k}}{k!} \exp \big( - \pi \lambda r^{2} \eta (s,m,r) \big) \\
& \phantom{=} \ \times \! B_{k} \bigg( \! - \pi \lambda r^{2} \frac{\diff}{\diff s} \eta (s,m,r), - \pi \lambda r^{2} \frac{\diff^{2}}{\diff s^{2}} \eta (s,m,r), \ldots, - \pi \lambda r^{2} \frac{\diff^{k}}{\diff s^{k}} \eta (s,m,r) \bigg) \bigg]_{s = m \theta r^{\alpha}} \! r \diff r \\
\nonumber & = 2 \pi \lambda \frac{1}{k!} \int_{0}^{\infty} \exp \big( - \pi \lambda r^{2} \eta (m \theta, m, 1) \big) \\
& \phantom{=} \ \times B_{k} \big( - \pi \lambda r^{2} \psi_{1} (\theta,m), - \pi \lambda r^{2} \psi_{2} (\theta,m), \ldots, - \pi \lambda r^{2} \psi_{k} (\theta,m) \big) r \diff r \\
& = \frac{1}{\eta (m \theta, m, 1)} \sum_{j=1}^{k} \frac{j!}{k!} B_{k,j} \bigg( \frac{\psi_{1} (\theta,m)}{\eta (m \theta, m, 1)}, \frac{\psi_{2} (\theta,m)}{\eta (m \theta, m, 1)}, \ldots, \frac{\psi_{k-j+1} (\theta,m)}{\eta (m \theta, m, 1)} \bigg)
\end{align}
with $\psi_{k}(\theta,m)$ defined in \eqref{eq:psi}. Finally, since \eqref{eq:P_cov_nlos_C1} for $k=0$ is equal to $\frac{1}{\eta (m \theta, m, 1)}$, we obtain $\Pcov^{\los,(\rmC)} (\theta,m)$ in \eqref{eq:P_cov_los_C} from \eqref{eq:P_cov_los}.

\textit{(b)} Let us now consider strongest BS association and let us recall the definition of $\setL_{I}^{\los,(\rmS)} (s)$ in \eqref{eq:L_I_los_S}. Since
\begin{align}
\frac{\diff^{k}}{\diff s^{k}} s^{\frac{2}{\alpha}} = (-1)^{k} \big( - \tfrac{2}{\alpha} \big)_{k} s^{\frac{2}{\alpha}-k}
\end{align}
for a given $k \geq 1$, we can derive the $k$th term of the summation in \eqref{eq:P_cov_los} as
\begin{align}
2 \pi \lambda \int_{0}^{\infty} & \bigg[ \frac{(-s)^{k}}{k!} \frac{\diff^{k}}{\diff s^{k}} \setL_{I}^{\los,(\rmS)} (s) \bigg]_{s = m \theta r^{\alpha}} r \diff r \\
\nonumber & = 2 \pi \lambda \int_{0}^{\infty} \bigg[ \frac{(-s)^{k}}{k!} \setL_{I}^{\los,(\rmS)} (s) \\
& \phantom{=} \ \times B_{k} \bigg( - 2 \pi \lambda \frac{\zeta(m)}{m^{\frac{2}{\alpha}}} \frac{\diff}{\diff s} s^{\frac{2}{\alpha}}, - 2 \pi \lambda \frac{\zeta(m)}{m^{\frac{2}{\alpha}}} \frac{\diff^{2}}{\diff s^{2}} s^{\frac{2}{\alpha}}, \ldots, - 2 \pi \lambda \frac{\zeta(m)}{m^{\frac{2}{\alpha}}} \frac{\diff^{k}}{\diff s^{k}} s^{\frac{2}{\alpha}} \bigg) \bigg]_{s = m \theta r^{\alpha}} r \diff r \\
\nonumber & = 2 \pi \lambda \frac{1}{k!} \int_{0}^{\infty} \setL_{I}^{\los,(\rmS)} (m \theta r^{\alpha}) \\
\label{eq:P_cov_nlos_S1} & \phantom{=} \ \times B_{k} \big( - 2 \pi \lambda r^{2} \phi_{1}(\theta,m), - 2 \pi \lambda r^{2} \phi_{2}(\theta,m), \ldots, - 2 \pi \lambda r^{2} \phi_{k}(\theta,m) \big) r \diff r \\
& = \frac{1}{2 \zeta(m) \theta^{\frac{2}{\alpha}}} \sum_{j=1}^{k} \frac{j!}{k!} B_{k,j} \bigg( \frac{\phi_{1} (\theta,m)}{\zeta(m) \theta^{\frac{2}{\alpha}}}, \frac{\phi_{2} (\theta,m)}{\zeta(m) \theta^{\frac{2}{\alpha}}}, \ldots, \frac{\phi_{k-j+1} (\theta,m)}{\zeta(m) \theta^{\frac{2}{\alpha}}} \bigg) \\
& = \frac{1}{2 \zeta(m) \theta^{\frac{2}{\alpha}}} \sum_{j=1}^{k} \frac{j!}{k!} B_{k,j} \Big( - \big( - \tfrac{2}{\alpha} \big)_{1}, - \big( - \tfrac{2}{\alpha} \big)_{2}, \ldots, - \big( - \tfrac{2}{\alpha} \big)_{k-j+1} \Big) \\
& = \frac{2}{\alpha} \frac{\zeta(k)}{k \zeta(m)} \Pcov^{\nlos,(\rmS)} (\theta)
\end{align}
where in \eqref{eq:P_cov_nlos_S1} we have introduced
\begin{align}
\phi_{k}(\theta,m) \triangleq - \big( - \tfrac{2}{\alpha} \big)_{k} \zeta(m) \theta^{\frac{2}{\alpha}}.
\end{align}
and with $\Pcov^{\nlos,(\rmS)} (\theta)$ defined in \eqref{eq:P_cov_nlos_S}. Finally, it is not difficult to show that
\begin{align}
\sum_{k=0}^{m-1} \frac{\zeta(k)}{k \zeta(m)} = \frac{\alpha}{2}
\end{align}
and we thus obtain $\Pcov^{\los,(\rmS)} (\theta,m) = \Pcov^{\nlos,(\rmS)} (\theta)$ from \eqref{eq:P_cov_los}. \hfill \IEEEQED

\subsection{Proof of Corollary~\ref{cor:P_cov_los}} \label{sec:A1_P_cov_los2}

\textit{(a)} Building on \cite[Eq.~9.137(2)]{Gra07}, we have
\begin{align}
\eta(2 \theta,2,1) = \frac{\big( 1 + \frac{2}{\alpha}) (1+\theta) \eta(\theta,1,1) - \frac{2}{\alpha}}{1+\theta}
\end{align}
which allows us to write
\begin{align}
\Pcov^{\los,(\rmC)} (\theta,2) & = \frac{1}{\eta(2 \theta,2,1)} \bigg( 1 + \frac{\frac{2}{\alpha} \big( \eta(2 \theta,2,1) + (1+\theta)^{-2} \big)}{\eta(2 \theta,2,1)} \bigg) \\
& = \frac{\big( 1 + \frac{2}{\alpha} \big) (1+\theta) \big( \big( 1 + \frac{2}{\alpha} \big) (1+\theta) \eta(\theta,1,1) - \frac{2}{\alpha} \big) - \frac{2}{\alpha}}{\big( \big( 1 + \frac{2}{\alpha} \big) (1+\theta) \eta(\theta,1,1) - \frac{2}{\alpha} \big)^{2}} \\
& > \Pcov^{\nlos,(\rmC)} (\theta).
\end{align}
The same property can be used recursively to show that $\Pcov^{\los,(\rmC)} (\theta,m+1) > \Pcov^{\los,(\rmC)} (\theta,m)$, $\forall m \geq 1$, with more involved calculations.

\textit{(b)} Building on Corollary~\ref{cor:P_cov_los}--\textit{(a)}, and since $\Pcov^{\los,(\rmC)} (\theta,m) \leq \Pcov^{\los,(\rmS)} (\theta,m)$, it follows that $\lim_{m \to \infty} \Pcov^{\los, (\rmC)} (\theta,m) = \Pcov^{\los, (\rmS)} (\theta,m)$, where $\Pcov^{\los, (\rmS)} (\theta,m) = \Pcov^{\nlos, (\rmS)} (\theta)$ (as derived in Theorem~\ref{th:P_cov_los}--\textit{(b)}). \hfill \IEEEQED

\section{The Effect of BS Height} \label{sec:A2}
\subsection{Proof of Lemma~\ref{lem:int}} \label{sec:A2_int}

Assume that the points of $\Phi$ are indexed such that their distances from the typical UE are in increasing order, i.e., $r_{x_{i}} \leq r_{x_{i+1}}$, $\forall i = 1, \ldots, \infty$. For strongest BS association, the expected interference power is given by
\begin{align}
\hspace{-2mm} \Exp \bigg[ \sum_{y \in \Phi \backslash \{x\}} g_{y} \ell(r_{y},h) \bigg] & < \Exp \bigg[ \sum_{x_{i} \in \Phi} g_{x_{i}} \ell(r_{x_{i}},h) \bigg] \\
\label{eq:int1} & = \sum_{i=1}^{\infty} \Exp \big[ g_{x_{i}} \ell(r_{x_{i}},h) \big] \\
\label{eq:int2} & = \sum_{i=1}^{\infty} \Exp \big[ \ell(r_{x_{i}},h) \big] \\
\label{eq:int3} & = \sum_{i=1}^{\infty} \int_{0}^{\infty} (r^{2} + h^{2})^{-\frac{\alpha}{2}} f_{r_{x_{i}}}(r) \diff r
\end{align}
where \eqref{eq:int2} follows from $\Exp \big[ g_{y} \ell(r_{y},h) \big] = \Exp [g_{y}] \Exp \big[ \ell(r_{y},h) \big]$ with $\Exp [g_{y}] = 1$ and where $f_{r_{x_{i}}}(r)$ in \eqref{eq:int3} is the pdf of the distance between the typical UE and the $i$-th nearest BS \cite[Ch.~2.9]{Hae12}:
\begin{align}
f_{r_{x_{i}}}(r) \triangleq e^{- \pi \lambda r^{2}} \frac{2 (\pi \lambda r^{2})^{i}}{r \Gamma(i)}.
\end{align}
Solving \eqref{eq:int3} for generic $i$ and plugging the result into \eqref{eq:int2} gives the expression on the right-hand side of \eqref{eq:int}. On the other hand, solving \eqref{eq:int3} for $i=1$ yields the expected interference power from the nearest interfering BS in \eqref{eq:int_nearest}. Evidently, since the terms in the summation in \eqref{eq:int1} are strictly decreasing with $i$ and the dominant interference term \eqref{eq:int_nearest} is finite, then the aggregate interference power is also finite. \hfill \IEEEQED

\subsection{Proof of Theorem~\ref{th:height}} \label{sec:A2_height}

Consider the pathloss function $\ell(r_{x},h) = (r^{2} + h^{2})^{-\frac{\alpha}{2}}$ and recall the definition of $\eta(s, m, r)$ in \eqref{eq:eta}.

\textit{(a)} The coverage probability for closest BS association is derived as
\begin{align}
\Pcov^{(\rmC)} (\theta, \lambda) & = 2 \pi \lambda \int_{0}^{\infty} \exp \bigg( - 2 \pi \lambda \int_{r}^{\infty} \bigg( 1 - \frac{1}{1 + \theta \frac{\ell (t, h)}{\ell (r, h)}} \bigg) t \diff t \bigg) e^{- \pi \lambda r^{2}} r \diff r \\
\label{eq:P_cov_C1} & = 2 \pi \lambda \int_{0}^{\infty} \exp \bigg( - 2 \pi \lambda \int_{\sqrt{r^{2}+h^{2}}}^{\infty} \bigg( 1 - \frac{1}{1 + \theta (r^{2}+h^{2})^{\frac{\alpha}{2}} t_{h}^{-\alpha}} \bigg) t_{h} \diff t_{h} \bigg) e^{- \pi \lambda r^{2}} r \diff r \\
\label{eq:P_cov_C2} & = 2 \pi \lambda \int_{0}^{\infty} \exp \big( - \pi \lambda (r^{2} + h^{2}) \big( \eta (\theta, 1, 1) - 1 \big) \big) e^{- \pi \lambda r^{2}} r \diff r \\
\label{eq:P_cov_C3} & = 2 \pi \lambda \int_{0}^{\infty} \exp \big( - \pi \lambda r^{2} \eta (\theta, 1, 1) \big) r \diff r \exp \big( - \pi \lambda h^{2} \big( \eta(\theta, 1, 1) - 1 \big) \big)
\end{align}
where in \eqref{eq:P_cov_C1} we have substituted $\sqrt{t^{2} + h^{2}} \to t_{h}$ in the inner integral. Finally, solving the integral in \eqref{eq:P_cov_C3} yields the expression of $\Pcov^{(\rmC)} (\theta, \lambda)$ in \eqref{eq:P_cov_C}. 

\textit{(b)} The coverage probability for strongest BS association is derived as
\begin{align}
\Pcov^{(\rmS)} (\theta, \lambda) & = 2 \pi \lambda \int_{0}^{\infty} \exp \bigg( - 2 \pi \lambda \int_{0}^{\infty} \bigg( 1 - \frac{1}{1 + \theta \frac{\ell (t, h)}{\ell (r, h)}} \bigg) t \diff t \bigg) r \diff r \\
\label{eq:P_cov_S1} & = 2 \pi \lambda \int_{h}^{\infty} \exp \bigg( - 2 \pi \lambda \int_{h}^{\infty} \bigg( 1 - \frac{1}{1 + \theta r_{h}^{\alpha} t_{h}^{-\alpha}} \bigg) t_{h} \diff t_{h} \bigg) r_{h} \diff r_{h}
\end{align}
where in \eqref{eq:P_cov_S1} we have substituted $\sqrt{t^{2} + h^{2}} \to t_{h}$ in the inner integral and $\sqrt{r^{2} + h^{2}} \to r_{h}$ in the outer integral. Finally, solving the inner integral in \eqref{eq:P_cov_S1} yields the expression of $\Pcov^{(\rmS)} (\theta, \lambda)$ in \eqref{eq:P_cov_S}. \hfill \IEEEQED

\subsection{Proof of Corollary~\ref{cor:height_limits}} \label{sec:A2_limits}

First, \textit{(a)} can be easily obtained from Theorem~\ref{th:height}--\textit{(a)}. Furthermore, \textit{(b)} is a consequence of Lemma~\ref{lem:L_I_0}. Lastly, \textit{(c)} follows from 
\begin{align}
\Pcov^{(\rmS)} (\theta, \lambda) & < 2 \pi \lambda \int_{0}^{\infty} \exp \bigg( - 2 \pi \lambda \int_{r}^{\infty} \bigg( 1 - \frac{1}{1 + \theta \frac{\ell (t, h)}{\ell (r, h)}} \bigg) t \diff t \bigg) r \diff r \\
& = 2 \pi \lambda \int_{0}^{\infty} \exp \big( - \pi \lambda (r^{2} + h^{2}) \big( \eta (\theta, 1, 1) - 1 \big) \big) r \diff r \\
\label{eq:P_cov_S2} & = \frac{1}{\eta(\theta,1,1) - 1} \exp \big( - \pi \lambda h^{2} \big( \eta(\theta, 1, 1) - 1 \big) \big)
\end{align}
and, since \eqref{eq:P_cov_S1} decays to zero as $\lambda \to \infty$, so does $\Pcov^{(\rmS)} (\theta, \lambda)$ (see Appendix~\ref{sec:A2_height} for details). \hfill \IEEEQED

\addcontentsline{toc}{chapter}{References}
\bibliographystyle{IEEEtran}
\bibliography{IEEEabrv,ref_Huawei}

\end{document}